\documentclass[prb,twocolumn,showpacs,showkeys,
               preprintnumbers,amsmath,amssymb,aps11]{revtex4}

\usepackage{graphicx}% Include figure files
\usepackage{dcolumn}% Align table columns on decimal point
\usepackage{bm}% bold math
\begin{document}

\title{ Maximally localized Wannier functions from PAW or ultrasoft pseudopotentials }

%%%%%%%%
%%%%%%%%
\author{Andrea \surname{Ferretti}}
\email[corresponding author: ]{ferretti.andrea@unimore.it}
\author{Arrigo \surname{Calzolari}}
\author{Benedetta \surname{Bonferroni}}
\author{Rosa \surname{Di Felice}}

\affiliation{National Research Center on nanoStructures
             and bioSystems at Surfaces ($S^3$), INFM-CNR
             and Dipartimento di Fisica, Universit\`a di
             Modena e Reggio Emilia, 41100 Modena, Italy}
%%%%%%%
%%%%%%%
\date{\today}

%*******************************
%  ABSTRACT
   %*******************************
\begin{abstract}
\noindent
We report a theoretical scheme that enables the calculation of maximally localized
Wannier functions in the formalism of projector-augmented-waves (PAW) which also
includes the ultrasoft-pseudopotential (USPP) approach.
We give a description of the basic underlying formalism
and explicitly write all the required matrix elements
from the common ingredients of the PAW/USPP theory.
We report an implementation of the method in a form suitable to accept the input
electronic structure from USPP plane-wave DFT simulations.
We apply the method to the
calculation of Wannier functions, dipole moments and spontaneous polarizations
in a range of test cases. 
Comparison with norm-conserving pseudopotentials is reported 
as a benchmark.
\end{abstract}
%\pacs{}

\maketitle

%*******************************
%  INTRO
   %*******************************
\section{Introduction}
Wannier functions (WFs) can be obtained by a unitary
transformation of the extended wavefunctions of a periodic
system.~\cite{wann37pr,kohn59pr} So far, the effective
diffusion/application of WFs  in electronic structure calculations
was hindered by their intrinsic
non-uniqueness.~\cite{wann37pr,kohn59pr} In 1997 Marzari and
Vanderbilt proposed a useful approach~\cite{marz-vand97prb,souz+01prb} 
to overcome this drawback: the
proposed methodology allows one to extract, from a selected
manifold of bands, the set of WFs with the maximum spatial
localization, {\it i.e.} the {\it maximally localized Wannier functions}
(MLWFs). On one hand, MLWFs are attractive because they
constitute a complete and orthonormal basis set in the real space.
On the other hand, with respect to other numerical real-space
basis sets, they carry also the physical information of the
starting Bloch functions. Indeed, MLWFs may yield the chemical
view of molecular bond orbitals, and they can be exploited
for the computation of the spontaneous polarization in periodic
systems,\cite{rest94rmp,king-vand93prb,vand-king93prb} becoming
very popular to tackle these issues in advanced
materials.~\cite{souz+00prb,bern+01prb,nakh+03prb} In addition,
MLWFs have most recently been proposed to calculate the transport
properties of nano-size conductors connected to external
electrodes.\cite{calz+04prb,ferr+05prl,ferr+05prb,lee+05prl,thyg-jaco05prl}

The calculation of MLWFs was originally
implemented~\cite{marz-vand97prb} for a projection from Bloch
orbitals expanded on a plane-wave basis set in  Density Functional
Theory (DFT) calculations, within the norm-conserving
pseudopotential (NCPP) framework. 
Norm-conserving
pseudopotentials,~\cite{hama+79prl,bach+82prb} that allow to
neglect core electrons in the evaluation of physical observables,
are usually characterized by a high transferability of an element
to a variety of chemical environments. However, the norm
conservation in the core region is a strong constraint, that
affects the computational effort of the DFT calculations. As a
consequence, only a part of the periodic table elements results
numerically accessible: some chemical species, such as first-row
elements (e.g. C, N, O, F) and especially transition metals (e.g. Mn, Fe, Co, Ni,
Cu) and rare earths (e.g. La, Gd, Yb), would require an extremely
high number of basis functions ({\it e.g.} plane waves), in order to be
described with a satisfactory accuracy. Unfortunally, the typical
systems of interest in nanoscience and specifically in molecular
electronics contain atoms of those critical species.

This problem was brilliantly solved within a frozen-core approach
by introducing the ultrasoft pseudopotentials
(USPPs),\cite{vand90prb} that relax the norm conservation
constraint, by compensating with a pseudized charge. Since the
computational requirement in the calculation of the DFT Bloch
functions  affects also the evaluation of the MLWFs, the
generalization of the original Marzari-Vanderbilt's procedure to
the case of USPPs becomes necessary, in order to  make the MLWFs a
powerful tool to tackle realistic systems of high technological
and fundamental relevance. This is the focal aim of the paper.

In pursuing this extension, we also noted that recent studies
mapped~\cite{hete+01jcp,kres-joub99prb} the USPP procedure into
the PAW theory. The latter~\cite{bloc90prb,bloc94prb} approach
was developed by Bl\"ochl to combine an all-electron description
of the system with the simplicity of the frozen-core
pseudopotential methods. The reader is referred to the original
articles for a description of the PAW scheme and for the matching
between PAW and USPP. 
The USPP can indeed be recast in the PAW formalism as an approximation, as we will
discuss in the following.
Given this equivalence, we developed a theoretical framework
within the PAW theory to compute WFs from USPP Bloch 
orbitals, as we show by expressing the necessary
matrix elements.

The paper is organized as follows: in section~\ref{sec:formalism}
we first write down the salient quantities of the USPP and PAW methods
that enter the computation of the WFs, and then obtain and discuss the
WF computation; in section~\ref{sec:applications} we report the
results of the application of our method to several test cases,
that explore both the chemical bonding and the electrical polarization in the pertinent
cases; finally we draw our conclusions in section~\ref{sec:conclusions}.

%*******************************
%  FORMALISM
   %*******************************
\section{Formalism}
\label{sec:formalism}

%-------------------
% ML Wannier functions
   %-------------------
\subsection{Maximally localized Wannier functions}
\label{sec:formalism_WF}
In this section we give a brief introduction to the theory of
maximally localized Wannier functions. A more detailed description
can be found in the original
papers.~\cite{marz-vand97prb,souz+01prb} In the case of an
isolated band, Wannier functions can be defined~\cite{wann37pr,kohn59pr}
as a combination of the Bloch orbitals $| \psi_{\mathbf{k}} \rangle$ 
corresponding to different $\mathbf{k}-$points as follows:
%
%
%*********** definition 1 band ************
\begin{equation}
\label{eq:wannier_definition_1band}
| w_{\mathbf{R}} \rangle =
\frac{V}{\left( 2 \pi \right)^3} \, \int_{BZ} \, d\mathbf{k}
        \, e^{-i \mathbf{k} \cdot \mathbf{R}} \, e^{i\phi_{\mathbf{k}}} \,
        | \psi_{\mathbf{k}} \rangle
\end{equation}
where $e^{i\phi_{\mathbf{k}}}$ is a $\mathbf{k}$-dependent phase factor.
This definition has been
generalized~\cite{marz-vand97prb} to a group of bands leading to
the expression:
%
%
%*********** definition  multiband ************
\begin{equation}
\label{eq:wannier_definition}
| w_{\mathbf{R},m} \rangle =
\frac{V}{\left( 2 \pi \right)^3} \, \int_{BZ} \, d\mathbf{k}
        \, e^{-i \mathbf{k} \cdot \mathbf{R}}
        \sum_n \, U^{\mathbf{k}}_{nm} | \psi_{\mathbf{k},n}
        \rangle \,\,.
\end{equation}
Here the extra degrees of freedom related to the phases of the
Bloch eigenstates are collected in the unitary matrix
$U^{\mathbf{k}}$. In the one-band case it has been demonstrated
that a suitable choice of the phases $e^{i\phi_{\mathbf{k}}}$ leads to WFs
which are real and exponentially decaying in a real space
representation.~\cite{kohn59pr} In the many-band
case~\cite{he-vand01prl} this theorem does not hold anymore but
the arbitrariness in the unitary (gauge) transformation
$U^{\mathbf{k}}$ can be exploited. Following Marzari and
Vanderbilt~\cite{marz-vand97prb} we define a {\it spread}
functional $\Omega\left[ U^{\mathbf{k}} \right]$, which gives a
measure of the degree of localization of the WF set. It reads:
%
%
%*********** spread ************
\begin{equation}
\label{eq:spread_functional}
      \Omega \left[ U^{\mathbf{k}} \right]=
      \sum_n \, \left[ \, \langle \hat{r}^2 \rangle_n - \langle \hat{\mathbf{r}}
      \rangle_n^2 \, \right] \, ,
\end{equation}
where $\langle \cdot \rangle_n$ is the expectation value of a
given operator on the $n$-th WF calculated using the
$U^{\mathbf{k}}$ gauge transformation. It is therefore possible to
define the {\it maximally localized} Wannier functions (MLWFs) as
the WFs resulting from Eq.~(\ref{eq:wannier_definition}) by means of the
unitary transformation $U^{\mathbf{k}}$ that minimizes the spread functional.

According to the formal analysis of the 
$\langle \hat{\mathbf{r}} \rangle_n$ and $\langle
\hat{r}^2 \rangle_n$ terms given by Blount,~\cite{blou62book,marz-vand97prb} it is
possible to demonstrate that the dependence of $\Omega$ 
on the gauge transformation is determined only by the
so-called overlap integrals $M^{\mathbf{k},\mathbf{b}}$:
%
%
%*********** overlaps ************
\begin{eqnarray}
\label{eq:overlap}
 \nonumber
             M^{\mathbf{k},\mathbf{b}}_{mn} &=&
             \langle \psi_{\mathbf{k},m} |
             \, e^{-i \mathbf{b} \cdot \hat{\mathbf{r}} }\,
             | \psi_{\mathbf{k}+\mathbf{b},n} \rangle  \\
             &=& \int_\text{Crystal} d\mathbf{r} \, u_{\mathbf{k},m}^{*}
             (\mathbf{r}) \, u_{\mathbf{k}+\mathbf{b},n}
             (\mathbf{r}),
\end{eqnarray}
$u_{\mathbf{k},m}(\mathbf{r})$ being the periodic part of the
Bloch states $\psi_{\mathbf{k},m} (\mathbf{r}) = e^{i \mathbf{k}
\cdot \mathbf{r}}\, u_{\mathbf{k},m} (\mathbf{r})$. The detailed
form of the position expectation values and of the spread
functional (including its gradient wrt to $U^{\mathbf{k}}$) in
terms of the overlap integrals is reported in
Appendix~\ref{app:wannier}. Since the representation of Bloch
eigenstates enters only the calculation of the
$M^{\mathbf{k},\mathbf{b}}$ integrals, these quantities are the
main objects to deal with when using a USPP or PAW formalism. 
The detailed treatment is reported in section \ref{sec:wannier_USPP}.

%-------------------
% PAW and USPP
   %-------------------
\subsection{PAW and ultra-soft pseudopotentials}
\label{sec:formalism_USPP}
The PAW formalism has been introduced by
Bl\"ochl~\cite{bloc90prb,vand-bloc93prb,bloc94prb} and it has also been
demonstrated~\cite{kres-joub99prb} that the Vanderbilt ultrasoft
pseudopotential (USPP)
theory~\cite{vand90prb,laas+91prb,laas+93prb,hete+01jcp,gian+04jcp}
can be obtained within the PAW approach.

Bl\"ochl's starting point is to partition the volume of the system
by setting spherical regions (atomic spheres) around each atom. In
each sphere two complete sets of wavelets~\cite{note_wavelet} ($\{
| \phi_i^{ae} \rangle \}$ and $\{ | \phi_i^{ps} \rangle \}$ )
localized in the sphere are defined. While the former, when
truncated, is intended to work with all-electron (AE) functions,
the latter should be smoother and easily representable in plane
waves. It is therefore possible to introduce a well defined linear
operator mapping one-to-one AE-like functions into smoother pseudo
(PS) functions and viceversa:
%
%
%*********** T_operation ************
\begin{equation}
\label{eq:T_operation}
      | \psi^{ae} \rangle = \hat{T} \, | \psi^{ps} \rangle \, ,
\end{equation}
where
%
%
%*********** T_definition ************
\begin{equation}
\label{eq:T_definition}
      \hat{T} = \sum_{I,i} \left( \, | \phi_{Ii}^{ae} \rangle -
                                           | \phi_{Ii}^{ps} \rangle \, \right)
                                           \langle \, \beta_{Ii}|
                                           \, .
\end{equation}
The index $I$ runs over different atoms ({\it e.g.} different
atomic spheres), while $ \langle \beta_{Ii} \, |$ are the
projectors~\cite{note_projectors} related to the pseudo wavelets
$| \phi_{Ii}^{ps} \rangle$. The $\hat{T}$ operator acts
on the pseudized functions to reconstruct~\cite{hete+01jcp} the AE ones.

Matrix elements and expectation values of a generic operator $\hat{A}$
on the physical AE states can be written as:
%
%
%*********** expectation value theory ************
\begin{equation}
\label{eq:expectation_value_theory}
    A_{mn}   %% = \langle \psi^{ae}| \hat{A} | \psi^{ae}\rangle
     = \langle \psi_m^{ps}| \hat{A}^{ps} | \psi_n^{ps}\rangle
     = \langle \psi_m^{ps}| \hat{T}^{\dagger} \hat{A} \hat{T} | \psi_n^{ps}\rangle \, .
\end{equation}
One of the main results of Ref.~[\onlinecite{bloc94prb}] is
the explicit expression for $\hat{A}^{ps}$, which we report here for the
case of local and semilocal operators:
%
%
%*********** expectation value ************
\begin{eqnarray}
\label{eq:expectation_value}
  \!\!\!\!\!\!\!\hat{A}^{ps} &=& \hat{A} + \hat{A}^{aug} \, , \\
\nonumber
  \!\!\!\!\!\!\!\hat{A}^{aug} &=&
     \sum_{I,ij} \left[
     \langle \phi_{Ii}^{ae} | \, \hat{A} \, | \phi_{Ij}^{ae} \rangle -
     \langle \phi_{Ii}^{ps} | \, \hat{A} \, | \phi_{Ij}^{ps} \rangle \right]
     \, | \beta_{Ii}^{} \rangle \langle \, \beta_{Ij}^{}| \, .
\end{eqnarray}
The second term ($\hat{A}^{aug}$) in the rhs of
Eq.~(\ref{eq:expectation_value}) takes into account the
corrections due to the use of the pseudo functions instead of the AE ones, and
from here on it will be called augmentation term.
At this point it is useful to define the quantities
$Q_{ij}^I(\mathbf{r})$ and $q_{ij}^I$ (augmentation densities and
charges respectively) as:
%
%
%*********** augmentation densities ************
\begin{eqnarray}
\label{eq:augmentation_densities}
Q_{ij}^I(\mathbf{r}) &=& \phi_{Ii}^{ae \, *}(\mathbf{r})\,
\phi_{Ij}^{ae}(\mathbf{r}) - \phi_{Ii}^{ps \, *}(\mathbf{r})\,
\phi_{Ij}^{ps}(\mathbf{r}) \, ,
\\
q_{ij}^I &=& \int d\mathbf{r} \, Q_{ij}^I(\mathbf{r}) \, .
\label{eq:augmentation_charges}
\end{eqnarray}
Within these definitions,
Eq.~(\ref{eq:expectation_value}) 
for local operators $A(\mathbf{r})$
can be recast in a more
convenient form:
%
%
%*********** expectation value local ************
\begin{equation}
\label{eq:expectation_value_local}
  \hat{A}^{aug} =
     \sum_{I,ij}
                 \left[ \int d\mathbf{r} \,
                 Q_{ij}^I(\mathbf{r}) \, A(\mathbf{r}) \right]
     \, | \beta_{Ii}^{} \rangle \langle \, \beta_{Ij}^{}| \, .
\end{equation}
Setting $\hat{A}$ to the identity in
Eqs.~(\ref{eq:expectation_value},\ref{eq:expectation_value_local}),
scalar products are given by $\langle \psi_m^{ps} | \, \hat{S} \,
| \psi_n^{ps} \rangle $ where the number operator~\cite{note_number_operator}
$\hat{S}$ (that characterizes also the USPP formalism~\cite{vand90prb}) is given by:
%
%
%*********** number operator ************
\begin{equation}
\label{eq:number_operator}
 \hat{S}  = \hat{T}^{\dagger}\hat{T} =
     \mathbb{I} +
     \sum_{I,ij}
     \, | \beta_{Ii}^{} \rangle \,q_{ij}^I \, \langle \,
     \beta_{Ij}^{}| \, .
\end{equation}
In the same way, by setting $A(\mathbf{r}') =
e\delta(\mathbf{r}'-\mathbf{r})$ we obtain an expression for the
density:
%
%
%*********** density ************
\begin{eqnarray}
\label{eq:density}
 \!\!\!\!n(\mathbf{r}) &=& n^{ps}(\mathbf{r}) + n^{aug}(\mathbf{r}) \, ,\\
     \nonumber
 \!\!\!\!n^{aug}(\mathbf{r}) &=& \frac{2e}{N_{\mathbf{k}}} \sum_{m\mathbf{k}} \sum_{I,ij}
                 \,\langle \psi_{\mathbf{k},m}^{ps} | \beta_{Ii}^{} \rangle \,
                 Q_{ij}^I(\mathbf{r}) \,
                 \langle \, \beta_{Ij}^{}| \psi_{\mathbf{k},m}^{ps} \rangle \, ,
\end{eqnarray}
where $n^{ps}(\mathbf{r})$ is the density contribution of the
pseudo wavefunctions.
Since we are able to express all the quantities of interest in
terms of the soft pseudo-states,
the quantum problem can be solved directly in this representation.
In order to do this, it
is necessary to {\it augment} the Hamiltonian operator: the procedure 
leads to additional terms which have exactly the same role as that of the
pseudopotentials in standard PW calculations.

Moving to the USPP framework, the generalization introduced 
by Vanderbilt~\cite{vand90prb} is twofold. 
(i) More than one projector per angular momentum channel 
can be taken into account: the inclusion 
of multiple projectors per channel enlarges the
energy range~\cite{bloc90prb,vand90prb} over which logarithmic
derivatives are comparable with the full potential case, thus
increasing the overall portability of the pseudopotential. 
(ii) By relaxing the
norm-conservation constraint of the pseudo reference-states, the pseudo wavefunctions 
are smoothened: thus, the required cutoff energy for PW representation
can be drastically lowered. 
The fact that properties (i) and (ii) are verified for the PAW 
wavelets $\phi_{i}$ establishes the connection between 
the PAW and USPP methods. 
In fact, (i) is naturally valid for 
the wavelets $\phi_{i}^{ps}$, because these form a basis set, 
and therefore have in
principle an infinite number of states for each angular momentum
channel. (ii) is valid as well, and the
possibility of non norm-conservation in passing from $|
\phi_{i}^{ae} \rangle$ to $| \phi_{i}^{ps} \rangle$ is accounted
by non null $q_{ij}^{I}$ terms in 
Eqs.~(\ref{eq:augmentation_densities},\ref{eq:augmentation_charges}). 
Consequently, the PAW theory for
wavefunction reconstruction can be basically adopted also in the case of
USPP.~\cite{vand-bloc93prb,kres-joub99prb}

While the PAW method is in principle an exact AE (frozen-core) approach,~\cite{note_PAW_frozen_core} 
the USPP method adopts a
further approximation~\cite{kres-joub99prb,note_paw_uspp} 
represented by the requirement of pseudizing the
augmentation densities $Q_{ij}^{I}(\mathbf{r})$
[Eq.~(\ref{eq:augmentation_densities})]. Since these terms contain
the AE reference states, they are not simply writable on a PW basis: 
the USPP pseudization is done to make them suitable for a PW representation.
This also means that the total density from
Eq.~(\ref{eq:density}) would be PW representable within such an
approach. Even though the augmentation densities are pseudized, they
must capture some features of the physical AE density.
Consequently, their PW cutoff energy may be larger than the one
associated to the pseudo wavefunction density
[$n^{ps}(\mathbf{r})$ in Eq.~(\ref{eq:density})].

%-------------------
% Wannier within PAW/USPP
   %-------------------
\subsection{Maximally localized Wannier functions within PAW/USPP}
\label{sec:wannier_USPP} As mentioned in
Sec.~\ref{sec:formalism_WF}, Bloch
wavefunctions enter the calculation of MLWF's only through the
overlap matrix elements $M^{\mathbf{k},\mathbf{b}}$, defined in
Eq.~(\ref{eq:overlap}). Therefore, the reconstruction of these
integrals from the knowledge of pseudo (ultrasoft) functions
completely solves the problem of computing MLWF's within 
a PAW/USPP formalism.
Once overlap matrices have been calculated, we do not longer distinguish
whether the parent Bloch wavefunctions were pseudized or not.

Being the overlaps $M^{\mathbf{k},\mathbf{b}}_{mn}$
[Eq.~(\ref{eq:overlap})] the matrix
elements of the local operator $e^{-i \mathbf{b} \cdot
\hat{\mathbf{r}} }$, we built up the corresponding augmentation using
Eq.~(\ref{eq:expectation_value_local}).
Overlaps can be written as:
%
%
%*********** overlap augmentation ************
\begin{eqnarray}
\label{eq:overlap_augmentation}
% pseudo term
             M^{\mathbf{k},\mathbf{b}}_{mn} &=&
             \langle u^{ps}_{\mathbf{k},m} |
             u^{ps}_{\mathbf{k}+\mathbf{b},n} \rangle + \\
% augmentation term
\nonumber
             &+& \sum_{I,ij} \,
                 Q_{ij}^I(\mathbf{b}) \,
                 \langle \psi_{\mathbf{k},m}^{ps} |
                 \beta_{Ii}^{\mathbf{k}} \rangle \,
                 \langle \, \beta_{Ij}^{\mathbf{k}+\mathbf{b}}|
                 \psi_{\mathbf{k}+\mathbf{b},m}^{ps} \rangle \, ,
\end{eqnarray}
where we have defined $Q_{ij}^I(\mathbf{b}) = \int d\mathbf{r} \,
Q_{ij}^I(\mathbf{r}) \,
                 e^{-i \mathbf{b} \cdot \mathbf{r} }$
and $| \beta_{Ii}^{\mathbf{k}} \rangle $ are $\mathbf{k}$-point
symmetrized projectors (Bloch sums).
Summation over ions $I$ is done in a single unit cell.
Details about the calculation of these quantities are reported in
Appendix~\ref{app:radial_ft}.
We stress that the scalar product of $| u^{ps} \rangle$ functions
cannot be simply augmented by the $\hat{S}$ [Eq.~(\ref{eq:number_operator})] operator
as those involving $| \psi^{ps} \rangle$'s.
In order to work with the periodic
part $|u^{ps}\rangle$ of the Bloch functions, {\it
first} we have to reconstruct the AE Bloch states by means of
$\hat{T}$ and {\it then} we can obtain the required
$|u^{ps}_{\mathbf{k},m}\rangle$ states by applying the local operator
$e^{-i\mathbf{k} \cdot \hat{\mathbf{r}} }$. Since this last
operator does not commute with $\hat{T}$, we are not allowed to
directly work on $|u^{ps}\rangle$ with the
reconstruction operator. 
In the first scalar product of Eq.~(\ref{eq:overlap_augmentation}) 
the number operator $\hat{S} =
\hat{T}^{\dagger} \hat{T}$ has been therefore substituted by the augmented
operator $\hat{M} = \hat{T}^{\dagger} \, e^{-i \mathbf{b} \cdot
\hat{\mathbf{r}} }\, \hat{T}$. 
We thus need to introduce the Fourier transform of
the augmentation densities $Q_{ij}^I(\mathbf{b})$ instead of the
augmentation charges $q_{ij}^I$. 
We also note that in the thermodynamic limit
the $\mathbf{k}$-point grid becomes a continuum, so that $\mathbf{b} \to
\mathbf{0}$. This limit leads to identical $\hat{S}$ and $\hat{M}$ operators. 
Therefore, within discrete $\mathbf{k}$-meshes the use
of $\hat{S}$ instead of $\hat{M}$ is an approximation expected to
give the best performance in the limit of a large number of
$\mathbf{k}$-points. We will refer to this approximation as the
thermodynamyc limit approximation (TLA). We will comment more on
its numerical aspects in Sec.~\ref{sec:applications}.

%
% Relation to the literature on the topic.
%
The problem of calculating Wannier functions within USPP has 
also been faced elsewhere in the
literature.~\cite{vand-king98condmat,bern-madd01jms,
marz+03psiknl} In a first attempt, Vanderbilt and
King-Smith~\cite{vand-king98condmat} extended the calculation of
the spontaneous polarization through the Berry
phase~\cite{king-vand93prb,vand-king93prb,rest94rmp} to the 
USPP procedure. Since the Berry phase~\cite{vand-king93prb} is directly
related to overlap integrals, an expression for its calculation is
reported in Eq.~(23) of Ref.~[\onlinecite{vand-king98condmat}].
This expression adopts the $\hat{S}$ number operator instead of
$\hat{M}$. Therefore, the result does not completely agree with the
one presented here [Eq.~(\ref{eq:overlap_augmentation})], but it
can be read as an approximating formula having the right
thermodynamic limit in view of the above discussion.
Bernasconi and Madden~\cite{bern-madd01jms,marz+03psiknl} derived
instead the formalism for MLWFs with USPP in a simplified
approach~\cite{silv+98ssc} valid only in the case of
$\Gamma$-sampled supercells. Although the basic ingredients ({\it
e.g.} overlaps) are the same, our treatment is valid for
generic periodic systems, and recovers the $\Gamma$-only
calculation as a special case. Furthermore, the proof given in
Ref.~[\onlinecite{bern-madd01jms}] is not based on a PAW
reconstruction as we did. The authors generalized the augmentation
operator for the density [Eq.~(\ref{eq:density})] also to the case
of the density matrix and then derived the augmentation for the
$e^{-i\mathbf{G}_i \cdot \hat{\mathbf{r}} }$ operator. Here
instead the $\mathbf{G}_i$ are the generators of the reciprocal
lattice.  We note that the result by Bernasconi and Madden is
coherent with our treatment, while the first one 
by Vanderbilt and King-Smith~\cite{vand-king98condmat} is not. 
Following Bernasconi and
Madden, Thygesen and coworkers~\cite{thyg+05prb} recently derived
an expression for overlap matrices within USPP for periodic
systems, considering a $\Gamma$-only supercell containing the
whole crystal.

Closing this section we wish to underline that matrix elements of the form
$ \langle \psi_{\mathbf{k},m} |
  \, e^{-i (\mathbf{q}+\mathbf{G}) \cdot \hat{\mathbf{r}} }\,
  | \psi_{\mathbf{k}+\mathbf{q},n} \rangle  $
enter also other physical problems. A particularly 
appealing case is 
the calculation of one-particle Green function in the GW 
approximation.~\cite{hedi65pr,onid+02rmp}
Current implementations~\cite{baro-rest86prb,hybe-loui86prb,godb+88prb} of the method
experience DFT wavefunctions mainly through the above defined matrix elements
(evaluation of the polarizability), while no direct access to the density is required.
The extension of GW calculations to the case of PAW~\cite{arna-alou01prb, lebe+03prb}
or USPP is therefore feasible
along the same lines we presented here for Wannier functions.
We note, however, that the numerical
cost is extremely higher due to the larger number of overlaps to be computed.

%----------------------------
%  IMPLEMENTATION
   %---------------------------
\subsection{Numerical details}
\label{sec:numerics}
In this section we would like to describe some issues related to the 
numerical performance
of the method. We implemented this formalism in the freely-available
\textsc{WanT} code,~\cite{WanT}
for the calculation of electronic and transport properties with WFs.
We also took advantage of the complete integration 
of the \textsc{WanT} code with the
\textsc{PWscf} package,~\cite{Pwscf} which explicitly treats 
the DFT problem using USPPs.
From here on we focus on the USPP formalism, 
even though a large part of the discussion is still
valid also in the PAW case.
The most important advantage of using the USPP construction for Wannier functions
is the scaling of the original DFT calculations, which has been described
elsewhere~\cite{laas+93prb} and we do not repeat now.
However, this scaling has also an effect on 
the actual computation of WFs and we analyze this aspect in detail.

As we described above, almost the whole amount of changes 
induced by the USPP description (relative to NCPP) in the
the calculation of MLWFs is related to the implementation of
Eq.~(\ref{eq:overlap_augmentation}). Details on how to compute $|
\beta_i^{\mathbf{k}} \rangle$ and $Q_{ij}(\mathbf{b})$ are
reported in Appendix~\ref{app:radial_ft}. Since no reference to
the charge is made, only the very smooth wavefunction grid is used
throughout the calculation. While a linear scaling with the number of
plane waves is exploited in the first term (pseudo overlaps) of
Eq.~(\ref{eq:overlap_augmentation}), scalar products between
projectors and pseudo states in the second term (augmentation
overlaps) are the price to pay for introducing USPPs. When we
consider that the number of $\beta$-projectors $N_{\beta}$ is of
the same order of the number of bands $N_b$ (but usually larger by
a factor between one and two) we see that both pseudo and
augmentation overlap terms have the same scaling, namely $N_b^2
\times N_{\mathbf{k}} \times N_{PW}$. However, the pseudo overlaps
turn out to have a larger prefactor~\cite{note_prefactor} and
represent the leading term. Usually, USPPs allow for a reduction of
the PW cutoff by a factor of 2 to 3 for first-row elements up to 5 or
even more for atoms with $d$- or $f$-states. This leads to a
reduction of the PW number $N_{PW}$ by a factor of around 3 to 10 or
more. Even if the scaling wrt $N_{PW}$ is linear, it more than
compensates the effort for augmenting overlaps and makes the
introduction of USPPs numerically advantageous. Our experience shows that
USPPs avoid the creation of bottlenecks in the computation of
overlaps and make the WF localization the leading part of the
calculation.

Finally, we note that in order to give a guess for the iterative minimizations
involved in the MLWF method, it is sometimes required to compute the projections of Bloch
states onto some starting localized functions.~\cite{marz-vand97prb,souz+01prb}
The augmentation of scalar products is
performed as usual accounting for the $\hat{S}$ number operator
[Eq.~\ref{eq:number_operator}]. Projections on the $\beta$-functions are 
required as well, but they have already been 
computed and it turned out that scaling is linear wrt the PW number as before.
Therefore, no implications on the above discussion arise.

%*******************************
%  APPLICATIONS
   %*******************************
\section{Applications}
\label{sec:applications}
In this Section we apply the above described formalism to some test cases, 
ranging from periodic crystals to isolated molecules. Precisely, we study
{\it fcc} Copper bulk, wurtzite AlN, and Watson-Crick DNA base pairs.
We address a number of physical properties connected to
Wannier functions: interpolation of the
electronic structure, calculation of dipole moments and spontaneous polarization,
analysis of the chemical bonding.
All the calculations are performed with both norm-conserving
and ultra-soft pseudopotentials. The numerical implications in using the USPP-TLA approach
for the augmentation of overlaps are also discussed.

%----------------------------
%  Copper Bulk
   %---------------------------
\subsection{Copper bulk}

We compute MLWFs for {\it fcc}-Copper, which has already
been used as a test case
in the literature~\cite{souz+01prb,marz+03psiknl,thyg+05prb} of WFs.
We adopt a $6 \times 6 \times 6$ mesh of $\mathbf{k}$-points
to sample the Brillouin zone and compute six WFs corresponding to the
lowest $s-d$ manyfold. In the disentangling procedure~\cite{souz+01prb}
(used to get the optimal subspace for WF localization)
we {\it freeze} the Bloch eigentates below the Fermi energy:
this means that the subspace
is constructed by these selected states plus a mixture of the states 
above the Fermi level.
We adopt a kinetic energy cutoff for wavefuntions of 120 Ry (25 Ry) when using
NCPP (USPP) and of 480 Ry (200 Ry) for the density.
\begin{figure}
      \includegraphics[clip,angle=-90,width=0.40\textwidth]{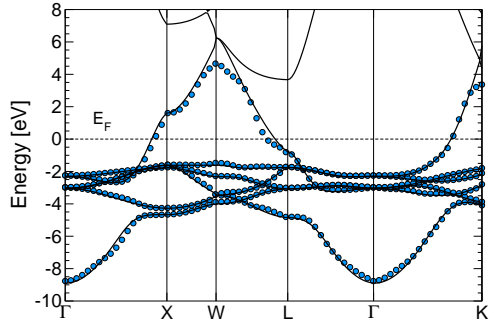}
      \caption{ \label{fig:copper-bands} (Color online)
      fcc{\it-Copper} band structure: solid lines represent DFT Kohn-Sham bands (USPP)
      while dotted lines are results calculated using WFs.
      }
   \end{figure}
In Fig.~\ref{fig:copper-bands} we superimpose the band structure
directly computed from a USPP-DFT calculation and that obtained from
Wannier function
interpolation~\cite{marz-vand97prb,souz+01prb,calz+04prb} on the 
adopted 8$\times$8$\times$8
uniform $\mathbf{k}$-point grid. These two sets of
bands are almost superimposed below the Fermi energy and some slight  
differences arise only at higher energies. This is expected due to
the choice of the energy window for the frozen states: while the
eigenvalues for the $\mathbf{k}$-points in the adopted regular
mesh are the same as those from the DFT calculation by
construction, it is not trivial that the band structure along a
generic Brillouin zone line is well reproduced. This is indeed the case here,
being a signature of proper localization of the computed WFs.
The band structure interpolation obtained from NCPP is essentialy the same
as the one in Fig.~\ref{fig:copper-bands} and it is not reported.
Some small descrepancies between NCPP and USPP interpolated bands are also 
present in the starting DFT calculations and are not of our interest in this context.

\begin{figure}[!b]
      \includegraphics[clip,width=0.38\textwidth]{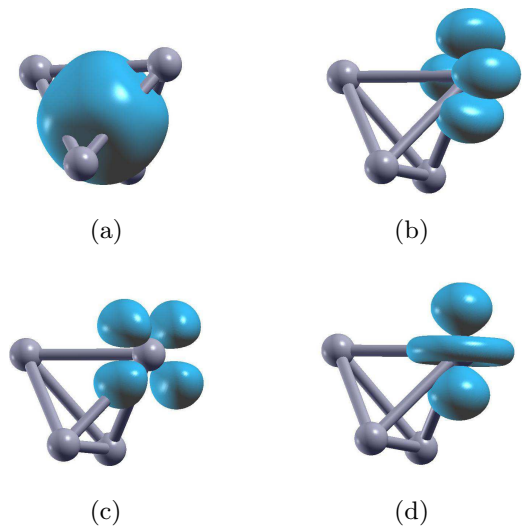}
      \caption{ \label{fig:copper-plot}
      fcc{\it-Copper}: charge distribution for WFs computed using USPP.
      (a) interstitial WF, $\Omega_a = 11.28$ Bohr$^2$; (b--d) $d$-character WFs,
      $\Omega_b = 1.86$ Bohr$^2$, $\Omega_c = 2.70$ Bohr$^2$, $\Omega_d = 1.54$ Bohr$^2$.
      The two $d$-like missing WFs are strictly similar to (b) and (c) and are not
      reported.
      }
   \end{figure}

In Tab.~\ref{tab:copper-data} we report a more detailed description of quantities
related to WFs (spreads, real-space decay of the Hamiltonian matrix elements)
in order to compare the NCPP and USPP approaches. 
A measure of the Hamiltonian decay is defined
as:
\begin{equation}
   \label{eq:decay}
   d(\mathbf{R}) =    \left( \, \frac{1}{N_w} \, \sum_{mn} \,
                      | H_{mn} (\mathbf{R}) |^2 \,\, \right)^{1/2},
\end{equation}
where we defined
$H_{mn} (\mathbf{R}) = \langle w_{\mathbf{0},m} | H |  w_{\mathbf{R},n} \rangle $.
\begin{table}
   \caption{fcc{\it-Copper}: WF spreads (Bohr$^2$) and real-space decay of the Hamiltonian
            matrix elements (eV) for NCPP, USPP and USPP in the thermodynamic limit
            approximation (USPP-TLA).
            $\Omega$ is the total spread, $\Omega_I$, $\Omega_D$ and $\Omega_{OD}$ are
            the invariant, diagonal and off-diagonal terms, according to
            Eq.~(\ref{eq:omega_tot}).
            $\boldsymbol{\tau}$ is the (0.5 0.5 0.0) direct lattice vector.
   \label{tab:copper-data}
   }
\begin{ruledtabular}
\begin{tabular}{p{1mm} l p{0mm} r p{0mm} r p{0mm} r}
 &            &&  \textbf{NCPP}  &&  \textbf{USPP}  &&  \textbf{USPP-TLA}  \\
\hline \\
 &  $\Omega$      &&  22.434     &&   23.010       &&    18.886         \\
 &  $\Omega_I$    &&  14.767     &&   15.629       &&    11.303         \\
 &  $\Omega_{D+OD}$    &&   7.667    &&    7.381       &&     7.583         \\
\\
 &  $d(\phantom{1}\boldsymbol{\tau})$  && 0.4876 &&    0.4795    &&   0.4779     \\
 &  $d(2\boldsymbol{\tau})$     &&        0.0493 &&    0.0473    &&   0.0476     \\
 &  $d(3\boldsymbol{\tau})$     &&        0.0203 &&    0.0207    &&   0.0205     \\

\end{tabular}
\end{ruledtabular}
\end{table}
USPP results appear to be in very good agreement with those related to NCPP, most
of the differences being reasonably due to the pseudopotential generation and not
to the WF computation. The average number of iterations to converge
the disentanglement and the localization procedures are almost the same, as well as
the singular spread values for each of the WFs.
Figure~\ref{fig:copper-plot} reports the spatial distribution of WFs from
USPP calculations. As in previous works,~\cite{souz+01prb,thyg+05prb}
we find one interstitial $s$-like WF with the largest spread 
[Fig.~\ref{fig:copper-plot}(a)], 
and five more localized $d$-like WFs [Fig.~\ref{fig:copper-plot}(b--d)], correctly 
reproducing the physical $s-d$ picture of Copper. 

As a last remark for the case of Copper, we analyze the effect of neglecting the
$e^{-i \, \mathbf{b} \cdot \mathbf{r} }$ term in Eq.~(\ref{eq:overlap_augmentation}),
{\it i.e.} the USPP thermodynamic limit approximation (USPP-TLA).
The theoretical background has already been discussed in
Sec.~\ref{sec:wannier_USPP}, here we focus on the numerical aspects.
In Tab.~\ref{tab:copper-data} the third column reports the resukts of 
the calculation performed
within this approximation: it is evident 
that the numerical values of the USPP-TLA spreads deviate from the NCPP 
ones much more than the USPP do.
On the contrary, the interpolated
band structure and the real-space decay Hamiltonian matrix elements are
definitely well-suited and comparable with those obtained in the
full USPP treatment.
Since the TLA is known to become exact in the thermodynamic limit, we expect
it to work better when incresing the dimension of the $\mathbf{k}$-point mesh.
We checked the behavior of the approximation with respect to different meshes but no
convergence could be reached for grids 
ranging from $4 \times 4 \times 4$ to 
$10 \times 10 \times 10$ $\mathbf{k}$-points.

%----------------------------
%  Molecules
   %---------------------------
\subsection{Isolated molecules: DNA bases and base pairs}
Wannier functions have been widely used to characterize the
electrostatic properties of several molecular systems ranging from
{\it e.g.} water~\cite{berg+00prb, leun-remp04jacs, dell-naor05cpc} and small
molecules~\cite{albe+99jpcb, mant+04jacs, kuo-tobi01jpcb, gaig-spri03jpcb} to large
biomolecules, such as proteins,\cite{sulp-carl00jpcb} nucleic
acids,~\cite{mund+02jpca,magi+06tobe} enzymes~\cite{cava-carl02jacs} and
ionic channels.~\cite{guid-carl02bba} In fact, the Wannier
transformation allows one to partition the charge density into
localized distributions of charges sitting on the
so-called Wannier centers $\langle \hat{\mathbf{r}}
\rangle_n$.\cite{rest94rmp,king-vand93prb} In the case of isolated
molecular systems, the dipole moment is a well defined quantity
given by $\mathbf{p} = \mathbf{p_{ion}} +\mathbf{p_{el}} \ $ with

\begin{eqnarray}
\label{eq:dipole}
\mathbf{p_{ion}}  &=&  +e\sum_I \, Z_I\mathbf{R}_I  \, , \\
\nonumber
\mathbf{p_{el}} &=& -2e \sum_n^{occ} \, \langle \hat{\mathbf{r}}
\rangle_n \, ,
\end{eqnarray}
where $\mathbf{p_{ion}}$ and $\mathbf{p_{el}} \ $ are the ionic
and electronic component respectively; $e$ is the electron charge;
the $I$ summation is over the ionic sites $\mathbf{R}_I$ and $Z_I$
is the valence charge of the $I^{th}$ atom, as defined by the
corresponding psudopotential; the $n$ summation is over the doubly
occupied valence states.

\begin{figure}
      \includegraphics[clip,width=0.40\textwidth]{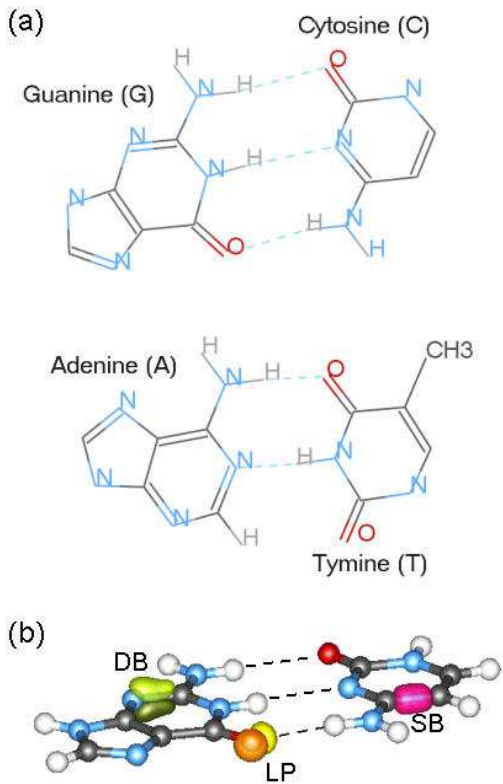}
      \caption{  (Color online) {\it DNA bases}:
      (a) Chemical scheme of the four DNA bases (G, C, A, T) assembled in
      the two Watson-Crick base pairs (GC and AT). (b) Isosurface plots of
      selected MLWFs for the GC base pair, representing a single C-C bond (SB),
      a double N=C bond
      (DB) and two oxygen lone pairs (LP). Dashed lines represent
      hydrogen bonds that bind the base pairs.
      }\label{fig:DNAWF}
   \end{figure}

\begin{table}[t!]
   \caption{{\it DNA bases}: Dipole moments $|\mathbf{p}|$ (Debye) for isolated
   nucleobases (G, C, A, T) and for Watson-Crick base pairs (G-C,
   A-T). Present work results for both  USPP and NCPP approaches are compared with
   previous quantum chemistry HF/6-31G$^{**}$ calculations~\cite{spon+96jpc} and
   experimental data.
   \label{tab:DNA-dipole}
   }
\begin{ruledtabular}
\begin{tabular}{p{1mm} l p{0mm} c p{0mm} c p{0mm} c p{0mm} c}
 &   &&  \textbf{USPP}  &&  \textbf{NCPP}  &&  \textbf{HF/6-31G$^{**}$} && \textbf{Exp.} \\
\hline \\
 &  G    &&  7.1     &&   7.2    &&   7.1   &&   7.1$^a$     \\
 &  C    &&  6.7     &&   6.9    &&   7.1   &&   7.0$^b$    \\
 &  A    &&  2.3     &&   2.3    &&   2.5   &&   2.5$^a$    \\
 &  T    &&  4.2     &&   4.4    &&   4.6   &&   4.1$^c$    \\
 &  G-C  &&  4.9     &&   4.9    &&   6.5   &&           \\
 &  A-T  &&  1.7     &&   1.8    &&   2.0   &&           \\
\end{tabular}
\end{ruledtabular}
\begin{flushleft}
$^{a}$DeVoe and Tinoco (Ref.~\onlinecite{devo-tino62jmb}) \\
$^{b}$Weber and Craven (Ref.~\onlinecite{webe-crav90acb}) \\
$^{c}$Kulakowski {\it et al.} (Ref.~\onlinecite{kula+74bba})
\end{flushleft}
\end{table}
As a key test, we calculated the dipole moments of the four
isolated DNA bases: Guanine (G), Cytosine (C), Adenine (A) and
Thymine (T), and of the two Watson-Crick base pairs G-C and A-T,
whose structures are reported in Fig.~\ref{fig:DNAWF}(a).

We simulated each system in a large ($ 22 \times 22 \times 22$)
\AA$^3$ supercell, which allows us to avoid spurious interactions
among neighbor replicas. For isolated systems, the uniform k-point
grid reduces to the case of $\Gamma$-point only, and the
connecting vectors $\mathbf{b}$ [Eq.~(\ref{eq:overlap})] correspond
to the generators of the reciprocal lattice vectors. We expanded
the electronic wavefuncions with the kinetic energy cutoff of 25
and 80 Ry, using USPP and NCPP respectively.

We first optimized the atomic structure of the two base pairs G-C
and A-T until forces on all atoms were lower than 0.03 eV/{\AA},
using ultra-soft pseudopotentials. Then, keeping atoms fixed in
the relaxed geometry, we calculated the electronic structure and
the corresponding MLWFs for both the single bases and the base
pairs. We maintained the same geometries also for the
corresponding NCPP calculations. In this case we have checked that
forces on single atoms never exceeded the value of 0.05eV/{\AA}.

Our results for both sets of calculations are reported in
Table~\ref{tab:DNA-dipole}. We note a very good internal agreement
between the USPP and NCPP cases, as well as in comparison with
previous quantum chemistry HF/6-31G$^{**}$
calculations~\cite{spon+96jpc} and experimental results. The total
spread $\Omega$ and its components $\Omega_I$, $\Omega_{OD}$ (not
shown) are also very similar in the two set of calculations, while
the diagonal term $\Omega_D$ is, by
definition,~\cite{marz-vand97prb} identically zero for
isolated systems.

Finally, we can take advantage from the calculated Wannier
functions to further investigate the electronic distribution and
the bonding pattern in the molecules.~\cite{fost-boys60rmp} For example,
as shown in Fig.~\ref{fig:DNAWF}(b) for the case of the G-C base pair,
we are able to characterize different kinds of bonds. The SB Wannier function
represents a single $\sigma$ {bond: It is}
centered in the middle point of the C-C bond. The two DB WFs
describe instead a double N=C
bond: This bond is partially polarized with a slight charge
accumulation near the nitrogen atom. Finally, the LP WFs
represents two electron lone
pairs localized around the oxygen atom of the guanine molecule.
Note also that the distribution of single/double bonds correctly
reflects the theoretical one reported in Fig.~\ref{fig:DNAWF}(a).
Let us remark here that our benchmark to assess the success of the newly 
developed USPP-WF 
methodology is its relative performance with respect to the NCPP-WF (already established) 
framework, namely the comparison between columns 2 and 3 in Table II.

%----------------------------
%  AlN 
   %---------------------------
\subsection{AlN Wurtzite}
Here we move to the calculation of polarization in
periodic systems. We focus on the case of aluminum nitride 
and compute the spontaneous 
polarization $P_{SP}$ of the wurtzite phase.
This is a particularly appealing test-case since a large debate exists in the literature
and many results are present.~\cite{bern-fior98prb,guy+99apl,rizz+99jvstb,
amba+99jap,bech+00prb,bern+01prb,amba+02jpcm} 
Moreover, Nitrogen may be more easily described 
(in terms of PW kinetic-energy cutoff) 
with USPPs than with NCPPs, allowing 
for an advantageous application of the current formalism. 
Following Refs.~[\onlinecite{bern+01prb,post+90prl}] we evaluate the polarization 
(electronic and ionic contributions)
for the wurtzite (WZ) phase taking the zinc-blend (ZB) structure as a reference.
Our calculations adopt the GGA-PBE parametrization for the 
exchange-correlation functional, and uses cutoff energies of 60 Ry (25 Ry) for NCPP (USPP) 
for wavefunctions and 240 Ry (200 Ry) for the density.
We relax both the cell dimensions and the atomic positions of the WZ phase. 
The ZB reference
is assumed to have ideal atomic positions and the cell is taken equal to the one
computed for the WZ polytype (six bilayers are included in the cell). 
These structural calculations have been performed using USPP and the obtained
(lattice and ionic) parameters have been used also in the NCPP simulations.
Using a $16\times 16\times4$ Monkhorst-Pack~\cite{monk-pack76prb} mesh of $\mathbf{k}$-points
we obtained the $a$ and $c/a$ parameters of the exagonal lattice as
$a$=3.1144 \AA{} and $c/a$=1.6109 (for the standard WZ cell including two bilayers, 
$c/a$=4.8326 for the actual cell we adopted).

In Tab.~\ref{tab:AlN-polarization} we report the comparison of
USPP and NCPP results in our calculations as well as other results from 
the literature. 
\begin{table}
   \caption{{\it AlN}: electronic spontaneous polarization [C/m$^2$] of wurtzite 
        structure. The presented results (NCPP, USPP and USPP-TLA) 
        are compared to the literature.
        As a reference we also report the invariant ($\Omega_I$) 
        and the total spread ($\Omega$) [Bohr$^2$] given by
        the WF calculations for wurtzite.
        Results from Ref.~[\onlinecite{bern+01prb}] are obtained using GGA and LDA
        (in parentheses) respectively.
       \label{tab:AlN-polarization}   
   }
\begin{ruledtabular}
\begin{tabular}{p{1mm} l p{0mm} c p{0mm} c p{0mm} c }
 &  P$_{SP}$&&  \textbf{NCPP}  &&  \textbf{USPP}  &&  \textbf{USPP-TLA} \\
\hline \\
 & this work        &&  -0.095  &&    -0.094         &&      -0.094      \\
 & Ref.~[\onlinecite{bern+01prb}]
       &&          &&                   && -0.090 (-0.099)  \\
 & Ref.~[\onlinecite{bech+00prb}]       
       &&          &&    -0.120         &&                  \\
\hline \\
 & $\Omega $        &&  80.895  &&    80.756         &&    80.271        \\
 & $\Omega_I $      &&  67.683  &&    67.857         &&    67.282        \\
\end{tabular}
\end{ruledtabular}
\end{table}
The values of spontaneous polarization computed using 
NCPPs are almost identical
to those obtained with USPPs ($P_{SP} =$-0.094 C/m$^2$). 
The USPP-TLA behaves very accurately 
in this case and no difference can be found with respect to the full USPP calculation.
Our results are also in very nice agreement with 
the USPP-TLA calculation by
Bernardini {\it et al.},~\cite{bern+01prb} 
who used a Berry-phase formalism~\cite{king-vand93prb} and 
found a value of $P_{SP} =$-0.090 C/m$^2$. We thus conclude 
that the USPP-TLA approximation performs well for the computation 
of the spontaneous polarization in nitrides, relative to a pure USPP treatment 
without the thermodynamic limit approximation.

These results for the P$_{SP}$ in AlN wurtzite are also in agreement
with indirect experimental evidencies as reported in
Refs.~[\onlinecite{amba+99jap,rizz+99jvstb,amba+02jpcm}].
While some earlier experimental fits~\cite{park-chua00apl,park+01jjap} 
claim for much lower values of P$_{SP}$ (ranging from -0.040 to -0.060 C/m$^-2$),
later~\cite{amba+02jpcm,vasc+02apl} works explain this discrepancy 
as due the neglecting of bowing effects
(non-linearity) of the P$_{SP}$ with respect to the composition of the
alloy Al$_{x}$Ga$_{1-x}$N employed in the measurements.

%*******************************
%        CONCLUSIONS
%*******************************
\section{Conclusions}
\label{sec:conclusions}
   In this paper we presented an approach to calculate {\it maximally localized
   Wannier functions} in the {\it ab initio} plane-wave ultrasoft pseudopotential 
   scheme. Our methodology is formulated in the general framework of the PAW theory
   and recovers the USPP framework as a special case.
   The main advantage using USPP is the well known reduction of
   the computational effort in the evaluation of the electronic
   wavefunctions at the DFT level. This leads to a consequent reduction 
   of computational load also in the calculation of MLWFs.
   We demonstrated that the extension to the USPP case
   does not introduce further approximations in the computation of the
   MLWFs with respect to the NCPP case. Furthermore, the
   reformulation within the PAW scheme allows us to interface
   the computation of MLWFs to other popular approaches for the
   electronic structure calculation.
   Finally, our method is formulated in the case of a uniform $\mathbf{k}$-point
   mesh for the sampling of the Brillouin Zone, generalizing
   previous attempts based on $\Gamma$-only calculations. This
   allows us to treat periodic solid-state systems (such as crystals,
   surfaces and interfaces), which require a full description of the BZ,
   as well as molecular, finite or amorphous systems which are
   well described with the $\Gamma$-only representation.

   As a first illustration of the capability of this methodology,
   we presented the calculation of the MLWFs for a few selected
   test cases, easily referable to well established theoretical and experimental
   results. For each selected system we also compared the results for
   both USPP and NCPP calculations, underlying a very good agreement between
   the two cases.

   The reduction of the computational cost resulting from USPP
   calculations opens the way to the exploitation of the MLWFS as
   a powerful tool to analyze the electronic structure
   of larger and more realistic nanoscale systems, in 
   particular for transport in nano-junctions.~\cite{calz+04prb,ferr+05prl}

%*******************************
%  ACKNOWLEDGMENTS
   %*******************************
\section{Acknowledgments}
We acknowledge discussions with Giovanni Bussi, Marco Buongiorno Nardelli
and Elisa Molinari for the treatment of USPP.
Funding was provided by the EC through 
TMR network ``Exciting'', by INFM through
``Commissione Calcolo Parallelo'', by the Italian MIUR through PRIN 2004,
and by the regional laboratory of EM ``Nanofaber''.
Part of the figures has been realized using the \textsc{XCrySDen} 
package.~\cite{koka03cms}

%*******************************
%  APPENDIX
   %*******************************
\appendix

%********************************
\section{Main formulas for maximally localized Wannier functions}
   %*******************************
   \label{app:wannier}
   For sake of completeness we report here the main relations \cite{marz-vand97prb}
   entering the expression~(\ref{eq:spread_functional}) for the
   spread functional $\Omega\left[ U\right]$ and its first derivative
   wrt the unitary transformation $U$ [Eq.~(\ref{eq:wannier_definition})].
   These are all the quantities involved in the minimization
   procedure for the calculation of maximally localized Wannier
   functions.
   The expectation values of the position operator are:
%
%
%*********** <rave> ************
\begin{eqnarray}
\!\!\!\!\! \langle \mathbf{r} \rangle_n = -\frac{1}{N}
\sum_{\mathbf{k},\mathbf{b}}
         w_b \mathbf{b} \, \text{Im} \, \text{ln}
         M^{\mathbf{k},\mathbf{b}}_{nn},
 \\
\!\!\!\!\! \langle r^2 \rangle_n = \frac{1}{N}
         \sum_{\mathbf{k},\mathbf{b}} w_b \, \{
         \left[ 1 - | M^{\mathbf{k},\mathbf{b}}_{nn}|^2 \right]
         + \left[ \text{Im} \, \text{ln}
         M^{\mathbf{k},\mathbf{b}}_{nn} \right]^2
         \},
\end{eqnarray}
where $\mathbf{b}$ vectors connect nearest-neighbor
$\mathbf{k}$-points and $w_b$ are their weights according to
Appendix B of Ref.~[\onlinecite{marz-vand97prb}].

The spread functional can be divided into three terms, the
invariant (I), the diagonal (D) and the off-diagonal (OD)
components:~\cite{marz-vand97prb}
%
%
%*********** spread decomposition ************
\begin{equation}
\label{eq:omega_tot}
\Omega [ U ] = \Omega_I + \Omega_D [U] + \Omega_{OD} [U],
\end{equation}
Their definitions are, respectively:
%
%
%*********** spread terms ************
\begin{eqnarray}
\label{eq:omegai}
\Omega_I &=& \frac{1}{N} \sum_{\mathbf{k},\mathbf{b}} w_b \,
             \left( N_w - \sum_{mn} |
             M^{\mathbf{k},\mathbf{b}}_{nn}|^2 \right), \\
\label{eq:omegad}
\Omega_D &=& \frac{1}{N} \sum_{\mathbf{k},\mathbf{b}} w_b \,
             \sum_{n} \left( \text{Im} \, \text{ln}
             M^{\mathbf{k},\mathbf{b}}_{nn} + \mathbf{b} \cdot
             \mathbf{r}_n \right)^2, \\
\label{eq:omegaod}
\Omega_{OD} &=& \frac{1}{N} \sum_{\mathbf{k},\mathbf{b}} w_b \,
             \sum_{m \neq n} |
             M^{\mathbf{k},\mathbf{b}}_{mn}|^2.
\end{eqnarray}

%********************************
\section{Detailed expressions for the calculation of $| \beta_{i}^{\mathbf{k}} \rangle$
         and $Q_{ij}(\mathbf{b})$ }
   %********************************
   \label{app:radial_ft}
We report the explicit expression for the calculation of the
reciprocal space representation of the PAW/USPP projectors and the
Fourier transform of the augmentation densities:
%
%
%*********** FT ************
\begin{equation}
    Q^I_{ij}(\mathbf{b}) = \int d\mathbf{r} \, e^{-i \mathbf{b} \cdot \mathbf{r}} \,
           Q^I_{ij}(\mathbf{r}-\boldsymbol{\tau}_I) \, .
\end{equation}
These tasks are also required in the evaluation {\it e.g.}
of the density in reciprocal space and
are therefore performed in standard plane-waves DFT codes.
The index $i,j$ in $\beta_i(\mathbf{r})$ and $Q_{ij}(\mathbf{r})$
stand for radial and angular numbers, {\it e.g.}
$n_i l_i m_i$. Projectors and augmentation densities are explicitly written
as a product of a radial part times (real) spherical harmonics:
%
%
%*********** augmentation expanded ************
\begin{eqnarray}
   \label{eq:augmentation_projectors_expanded}
   \beta_{i}(\mathbf{r}) &=& R_{n_i}(r) \, Y_{l_i}^{m_i}(\hat{r}) \,
\\
   \label{eq:augmentation_densities_expanded}
   Q_{ij}(\mathbf{r}) &=& g_{n_i n_j}(r) \, Y_{l_i}^{m_i}(\hat{r}) \,
                        Y_{l_j}^{m_j}(\hat{r}) .
\end{eqnarray}

First we focus on the expression for $\beta$ projectors.
Functions of the form $ f(\mathbf{r}) = R(r)\, Y_L^M(\hat{r})$ have a known
semi-analytical Fourier transform which is given by:
%
%
%*********** radial FT ************
\begin{equation}
   \label{eq:radial_ft}
   f(\mathbf{k}) = 4\pi \, (-i)^l \, Y_l^m(\hat{k}) \int_0^{\infty}
                   r^2 \, R(r) \, J_l(kr) dr
\end{equation}
where $J_l(x)$ is the spherical Bessel function of order $l$.
The problem for $\beta$ projectors is therefore directly solved once we add the
structure factors accounting for the acutal positions of the ion:
%
%
%*********** beta_k ************
\begin{multline}
   \label{eq:beta_k}
   \beta_{Ii}^{\mathbf{k}}(\mathbf{G}) =
              4\pi \,
              (-i)^{l_i} \,
              e^{-i(\mathbf{k}+\mathbf{G})\boldsymbol{\tau}} \,
              Y_{l_i}^{m_i}(\hat{G}) \,
              \times \\ \times
              \int_0^{\infty} r^2 \, R_{n_i}(r) \, J_{l_i}(Gr) dr  .
\end{multline}

Moving to the case of the augmentation densities, we note that
the product of two spherical harmonics can be expressed as a linear combination
of single spherical harmonics using Clebsch-Gordan coefficients:
%%
%%
%%*********** clebsch-gordan ************
%\begin{multline}
%   \label{eq:clebsch-gordan}
%   Y_{l_i}^{m_i}(\hat{r}) \, Y^{l_j}_{m_j}(\hat{r}) =
%         \sum_{l=|l_i-l_j|}^{l_i+l_j} \sum_{m=-l}^{l}
%         \left[ \frac{(2l_i+1)(2l_j+1)}{4\pi(2l+1)} \right]^{\frac{1}{2}} \, \\
%         \langle l_i l_j 00 | l0 \rangle \, \langle l_i l_j m_i m_j | l m \rangle \,
%         Y_{l}^{m}(\hat{r})
%\end{multline}
%%
%%
%
%
%*********** clebsch-gordan ************
\begin{multline}
   \label{eq:clebsch-gordan}
   Y_{l_i}^{m_i}(\hat{r}) \, Y_{l_j}^{m_j}(\hat{r}) =
         \sum_{L=|l_i-l_j|}^{l_i+l_j} \sum_{M=-L}^{L}
         C_{l_i l_j m_i m_j}^{LM} \,
         Y_{L}^{M}(\hat{r}) \, .
\end{multline}
This allows to follow the same strategy as before also for
Eq.~(\ref{eq:augmentation_densities_expanded}).
Putting Eqs.~(\ref{eq:clebsch-gordan}-\ref{eq:radial_ft}) together,
the final expression for $Q^I_{ij}(\mathbf{b})$ reads:
%
%
%*********** radial FT ************
\begin{multline}
   \label{eq:augmentation_densities_ft}
    Q^I_{ij}(\mathbf{b}) = 4\pi \, e^{-i\mathbf{b}\cdot \boldsymbol{\tau}} \,
          \sum_L \sum_M \,
          C_{l_i l_j m_i m_j}^{LM} \, (-i)^L \,
          \times \\ \times
          Y_L^M(\hat{b}) \,
          \int_0^{\infty} r^2 \, g_{n_i n_j}(r) \, J_L(br) dr \, .
\end{multline}
where $M,L$ indexes run as in Eq.~(\ref{eq:clebsch-gordan}).

%*******************************
%  BIBLIOGRAPHY
   %*******************************


\begin{thebibliography}{10}
\bibitem{note_projectors}
Since the pseudo wavelets are not orthogonal their projectors are not simply
  the associated bra's of the given ket's.
\bibitem{note_PAW_frozen_core}
PAW are presented here in a all-electron frozen-core fashion but a full
  all-electron implementation is also compatible with the
  theory.~\cite{bloc94prb}.
\bibitem{note_prefactor}
This fact comes first due to the presence of the sum over $\mathbf{b}$-vectors
  (giving a prefactor from 3 to 6) and secondly because the coupling of
  different $\mathbf{k}$-points for pseudo overlaps makes wavefunction managing
  more cumbersome.
\end{thebibliography}
\end{document}